\documentclass[amsmath,amssymb,nofootinbib,prl]{revtex4}
\pdfoutput=1
\usepackage{graphicx}
\usepackage{epsfig}
\usepackage{rotate}
\usepackage{amsmath}
\usepackage{amssymb}
\usepackage{amsfonts}
\usepackage{enumerate}
\usepackage{afterpage}
\usepackage{color}

\newcommand{\ud}{\mathrm{d}}
\newcommand{\p}{\partial}
\newcommand{\cH}{\mathcal{H}}
\newcommand{\Perp}{\mathcal{P}}

\def\be{\begin{equation}}
\def\ee{\end{equation}}
\def\bea{\begin{eqnarray}}
\def\eea{\end{eqnarray}}

\begin{document}

\title{Observed galaxy number counts on the lightcone up to second order: I. Main result}

\author{Daniele Bertacca$^{a}$, Roy Maartens$^{a,b}$, Chris Clarkson$^{c}$\\~}

\affiliation{
$^a$Physics Department, University of the Western
Cape, Cape Town 7535, South Africa\\
$^b$Institute of Cosmology \& Gravitation, University of
Portsmouth,
Portsmouth PO1 3FX, UK \\
$^c$Centre for Astrophysics, Cosmology \& Gravitation, and, Department of Mathematics \& Applied Mathematics, University of Cape Town, Cape Town 7701, South Africa}

\begin{abstract}
We present the galaxy number overdensity up to second order in redshift space on cosmological scales  for a concordance model. The result contains all general relativistic effects up to second order that arise from observing on the past light cone, including all redshift effects, lensing distortions from convergence and shear, and contributions from velocities, Sachs-Wolfe, integrated SW and time-delay terms. This result will be important for accurate calculation of the bias on estimates of non-Gaussianity and on precision parameter estimates, introduced by nonlinear projection effects. 

\end{abstract}

\date{\today}

\maketitle

\noindent{\bf Introduction}\\

The galaxy fractional number overdensity $\delta_g=\delta n_g/n_g$ at first order of perturbations is usually related to the matter fractional overdensity $\delta_m$ as \cite{Matsubara:2000pr}
\be \label{del1}
\delta_g=b\delta_m-{1\over {\cal H}}(n^i\partial_i)^2v-2\kappa,
\ee
where $b=b(z)$ is the galaxy bias, $\partial_iv$ is the galaxy peculiar velocity in the Kaiser redshift-space distortion term and $\kappa$ is the weak gravitational lensing integral. The Kaiser and lensing terms can be thought of as relativistic corrections to $\delta_g$ that are necessary on sub-Hubble scales ($\kappa$ is only significant at higher redshift). There are further relativistic effects that can be important on scales near and beyond the Hubble scale. On these scales,  $\delta_g$ is gauge dependent, which means that we have to construct the unique physical number overdensity that is observed on the lightcone, $\Delta_g$. This physical quantity is automatically gauge-invariant and can be computed in any chosen gauge. In Newtonian gauge this gives \cite{Yoo:2009au, Yoo:2010ni, Bonvin:2011bg, Challinor:2011bk,Jeong:2011as,Bertacca:2012tp}
\bea
\Delta_g &=& b\delta_{m{\rm S}}-{1\over {\cal H}}(n^i\partial_i)^2v-2\kappa \nonumber \\
&&{}+(3-b_e){\cal H}v+\left[b_e-{{\cal H}'\over {\cal H}^2}-{2\over \bar  \chi{\cal H}}\right]\left[ n^i\partial_i v -\Phi-2\int_0^{\bar \chi} d\tilde\chi\,\Phi' \right] 
- \Phi+{\Phi'\over {\cal H}}+{4\over \bar \chi}\int_0^{\bar \chi} d\tilde\chi\,\Phi, \label{delrel}
\eea
where $\delta_{m{\rm S}}$ is in synchronous-comoving gauge (to give the correct definition of bias on large scales), $b_e=3+d\ln n_g/d\ln a$ is the evolution bias, $\bar \chi$ is the comoving distance to the source and $\Phi$ is the metric perturbation, where the relativistic Poisson equation is $\nabla^2\Phi=(3/2){\cal H}^2\Omega_m\delta_{m{\rm S}}$. We have omitted terms evaluated at the observer. These terms are shown below in \eqref{Poiss-Deltag-1}. (We have also neglected magnification bias, leaving this for future work  \cite{Bertacca:2014hwa}.)

Here we give the second-order extension of \eqref{delrel}  on cosmological scales, including all general relativistic effects. The detailed derivation, which is in a general gauge and also includes general dark energy and modified gravity models, is given in an accompanying paper \cite{Bertacca:2014wga}. The second-order extension is relevant for an accurate calculation of the contamination of primordial non-Gaussianity on large scales by second-order projection effects \cite{Bertacca:2014n}.

{\em Note added in version 5:} In the previous version 4 (and in the published paper), we mistakenly omitted some terms which arise from the integration of a first-order quantity taking into account perturbations of the direction of the null geodesic (so-called post-Born terms). This error has been corrected here and in the companion paper 1406.0319v4. Our results are now in agreement, in the appropriate limit, with those of \cite{Nielsen:2016ldx}.

~\\
\noindent{\bf Second-order number counts on the lightcone}\\

We assume a concordance background and at first order we neglect anisotropic stress,
vector and tensor perturbations. 
In Poisson gauge, the metric and peculiar velocity are 
 \begin{eqnarray} 
 \label{Poiss-metric}
 \ud s^2 &=& a(\eta)^2\left\{-\left(1 + 2\Phi +\Phi^{(2)}\right)\ud\eta^2+2\omega_{i}^{(2)}\ud\eta \, \ud x^i+\left[\delta_{ij} \left(1 -2\Phi -\Psi^{(2)}\right)+\frac{1}{2}\hat h_{ij}^{(2)}\right]\ud x^i\ud x^j\right\} ,\\
v^{i  }&=& \p^i v +\frac{1}{2}v^{i (2)},~~ v^{i (2)}= \p^i v^{(2)}+ \hat v^{i (2)}, 
\end{eqnarray}
where we omit the superscript (1) on familiar quantities such as  $\Phi$ and $\p^iv$. At second order, the first-order scalars generate vector perturbations $\omega_{i}^{(2)}, \hat v^{i (2)}$ and a tensor perturbation $\hat h_{ij}^{(2)}$.
\begin{figure}[!htbp]
\centering
\includegraphics[width=4.5 cm]{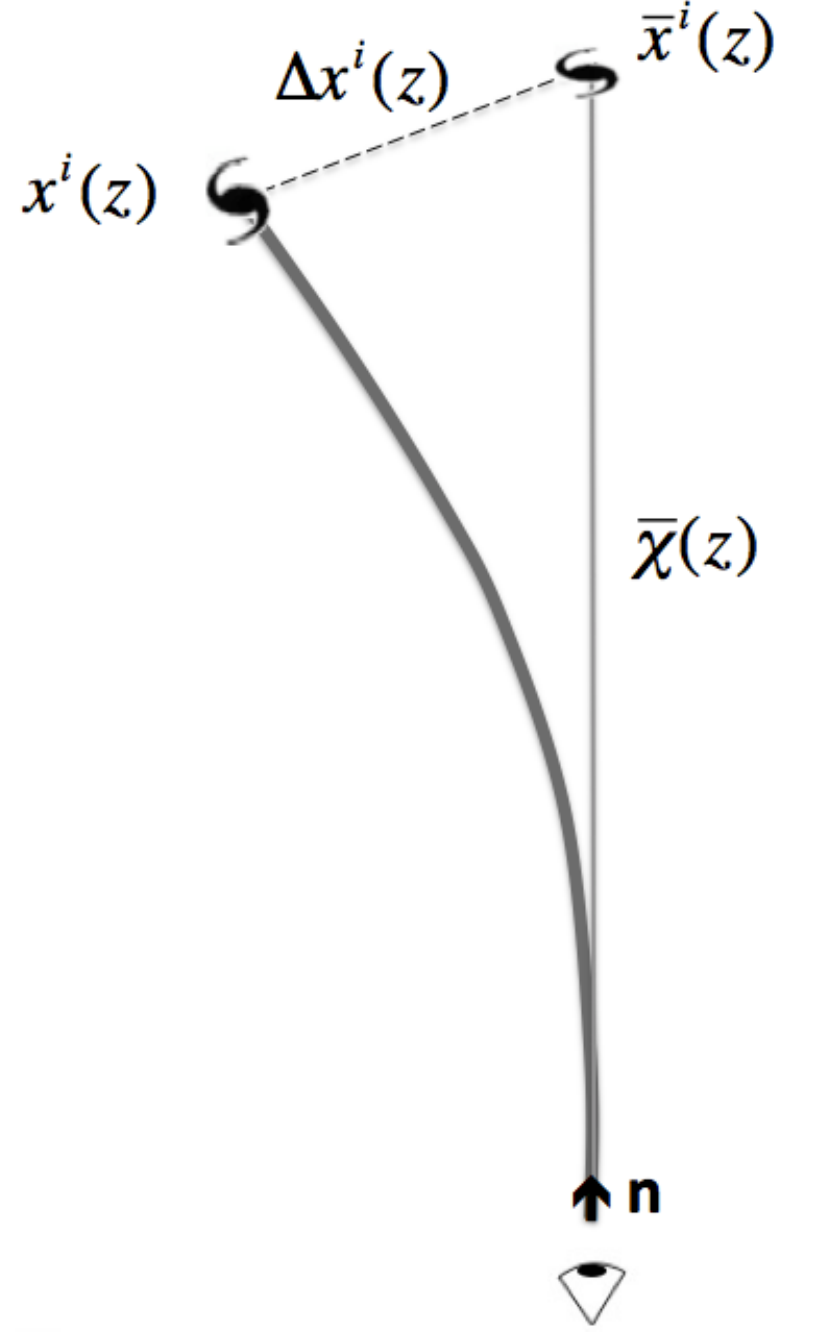}
\caption{The real-space and redshift-space views. 
\label{fig:1}
}
\end{figure}

We use only the observed redshift $z$ in our analysis. In particular, all background quantities are evaluated at the observed, not background, redshift. Thus we do not need to identify the perturbations of redshift (these are derived up to second-order by  \cite{Umeh:2012pn,Umeh:2014ana}). We set up a map between redshift space and real space (see Fig. \ref{fig:1}), generalizing the cosmic rulers approach of  \cite{Jeong:2011as,Schmidt:2012ne} from first to second order.
The observed galaxy has conformal coordinates $\bar{x}^\mu=(\bar \eta,\; \bar {\bf x})=(\eta_0-\bar \chi, \; \bar \chi \, n^i)$ in redshift space. The real space coordinates are $x^\mu(\chi)=\bar{x}^\mu(\bar\chi)+\Delta x^{\mu (1)} (\bar \chi)+\Delta x^{\mu (2)} (\bar \chi)/2$. 
%where $\bar \chi(z)$ is the zeroth order comoving distance in redshift space, $z$ is the redshift  and ${\bf n}$ is the direction to the observed galaxy, i.e. $n^i=\bar x^i/\bar \chi=\delta^{ij} (\p \bar \chi/\p \bar x^j)$. 
The spatial and temporal deviations encode information about volume and lensing distortions. Further details are given in \cite{Bertacca:2014wga}. For example, if we define
 \begin{eqnarray}
 \label{kappa-n}
\kappa^{(n)}\equiv- \frac{1}{2}  \p_{\perp i} \Delta x_{\perp}^{i (n)} ,
\end{eqnarray}
then at first order ($n=1$) we recover the lensing convergence integral $\kappa$: 
\begin{eqnarray}
 \label{Poiss-kappa-1}
 \kappa^{(1)} =\kappa- v_{\|  o}= \int_0^{\bar \chi} \ud \tilde \chi  \left(\bar \chi-\tilde \chi\right) \frac{\tilde \chi}{ \bar \chi} \,   \tilde \nabla^2_\perp \Phi  - v_{\|  o} , 
 \end{eqnarray}
with a Doppler correction at the observer. Here $\perp$ denotes projection into the screen space (with projector $\Perp^{ij}= \delta^{ij} -  n^i n^j$), $\|$ indicates projection along the unit line of sight vector $n^i$, and we define the derivatives 
\begin{eqnarray}
&&\partial_\parallel=n^j \partial_j,
%~~\partial_\parallel^i =n^i \partial_\parallel, ~~\partial^2_\parallel =
%\partial_{\parallel i}\partial_\parallel^i =\partial_\parallel
%\partial_\parallel, 
~~ \partial_\perp^i = \Perp^{ij}\partial_j= \partial^i -  n^i \partial_\parallel
,~~ \nabla^2_\perp = \partial_{\perp i}\partial_\perp^i =\nabla^2
- \partial_\parallel^2 - 2 {\chi}^{-1}\partial_\parallel.
\end{eqnarray}

At first order, we find the observed fractional number overdensity as
\begin{eqnarray}
\label{Poiss-Deltag-1}
\Delta_g 
 &=& \delta_g  + \left( b_e  - \frac{\cH'}{\cH^2} - \frac{2}{\bar \chi \cH}\right)\Delta \ln a^{(1)} - \frac{1}{\cH}  \p_\|^2 v + \frac{1}{\cH}  \Phi {'} - \Phi - \frac{2}{\bar \chi} T^{(1)} - 2 \kappa^{(1)} ,\\
 \Delta \ln a^{(1)}&=&{a\over \bar a}-1=\Phi _o-v _{\|  o} - \Phi + \p_\| v + 2I^{(1)},
\end{eqnarray}
which is in agreement with \eqref{delrel} since 
\bea \label{Poiss-iota}
T^{(1)}=- 2 \int_0^{\bar \chi} \ud \tilde \chi \Phi ,~~~I^{(1)}=   - \int_0^{\bar \chi} \ud \tilde \chi \, \Phi {'}.
\eea
$T^{(1)}$ is a radial displacement corresponding to the usual (Shapiro) time delay \cite{Challinor:2011bk}, and $I^{(1)} $ is the integrated Sachs-Wolfe (ISW) term. We also use 
\begin{eqnarray}
S^{i(1) } =-  \int_0^{\bar \chi} \ud \tilde \chi \left( \tilde\p^i \Phi   -\frac{1}{\tilde \chi} n^i\Phi  \right). 
\label{Poiss-varsigma}
\end{eqnarray}
%\begin{eqnarray}
%S_{\perp}^{i } =\Perp^i_j S_{\perp}^{j }= - \int_0^{\bar \chi} \ud \tilde \chi \, \tilde\p^i_\perp \Phi   \;, \quad \quad \quad S_{\|}  = n_i S^{i } =\Phi _o  -  \Phi + I  + \int_0^{\bar \chi} \ud \tilde \chi \frac{\Phi}{\tilde \chi} \;. 
%\end{eqnarray}

At second order we obtain \cite{Bertacca:2014wga}
\begin{eqnarray}
\label{Poiss-Deltag-2}
  \Delta_g^{(2)} &=&  \delta_g^{(2)} + \Phi^{(2)}  - 2  \Psi^{(2)} -\frac{1}{2}\hat h_{\|}^{(2)}+  \frac{1}{ \cH} \Psi^{(2)}{'}-  \frac{1}{2 \cH} \hat h^{(2)}_{\| }{'}   -\frac{1}{ \cH }\p_\|^2v^{(2)} -\frac{1}{ \cH }\p_\| \hat v_\|^{(2)}  +  \left( b_e  -   \frac{\cH'}{\cH^2} -\frac{2}{\bar \chi \cH}\right) \, \Delta \ln a^{(2)}  \nonumber \\
&& - \frac{2}{\bar \chi} T^{(2)}  - 2\kappa^{(2)} + \left(\Delta_g  \right)^2 - \left(\delta_g \right)^2 -14   {\Phi}^2+ \frac{1}{\cH^2}\left(\p_\|^2 v  \right)^2+\left( \p_\| v  \right)^2 +\frac{4}{\bar \chi  \cH} \left( \p_\| v  \right)^2+ \frac{1}{\cH^2}\left( \Phi {'}  \right)^2  
  +\frac{2}{\cH}\p_\| v  \Phi {'} \nonumber \\ 
&&  + \frac{2}{ \cH }\Phi  \Phi {'}+2\frac{\cH'}{\cH^3}\Phi \Phi {'}+\frac{4}{ \cH }\p_\| v  \p_\| \Phi  - \frac{4}{\cH} \Phi \p_\|^2 v   - \frac{2}{\cH^2} \Phi \p_\|^3 v   -\frac{2}{\cH}\Phi \p_\| \Phi + \frac{2}{\cH^2}\Phi \frac{\ud \,}{\ud \bar \chi}\Phi {'} - \frac{2}{\cH^2}\p_\| v \frac{\ud \,}{\ud \bar \chi}\Phi {'}   + \frac{2}{\cH^2} \p_\| v  \p_\|^2 \Phi \nonumber \\ 
&&    -2\frac{\cH'}{\cH^3}\Phi  \p_\|^2 v    +\frac{6}{\cH} \p_\| v \p_\|^2 v  +2\frac{\cH'}{\cH^3} \p_\| v \p_\|^2 v  - \frac{2}{\cH^2}\Phi  \p_\|^2 \Phi   -2\frac{\cH'}{\cH^3} \p_\| v \Phi {'}-  \frac{2}{\cH^2}\p_\|^2 v   \Phi {'}     +\frac{2}{\cH}\p_{\perp i} v \p^i_\perp \Phi    -\frac{2}{ \cH } \p_{\perp i} v   \p_{\perp}^i \p_\|  v \nonumber \\  
&&  +\frac{2}{ \bar \chi \cH } \p_{\perp i} v    \p_{\perp}^i v   - \p_{\perp i} v   \p_{\perp}^i v  +  \frac{2}{\cH^2}\p_\| v \p_\|^3v  +\frac{2}{ \cH } \p_\| v   \nabla^2_{\perp}  v  + \bigg( - \frac{8}{\bar\chi \cH}   \Phi  -4\frac{1}{\cH} \Phi {'}   - \frac{2}{\cH}\frac{\ud \,}{\ud  \bar \chi} \delta_g   -  \frac{4}{\cH \bar \chi^2}  T^{(1)} \nonumber \\  
&&  - \frac{4}{\bar \chi \cH}   \kappa^{(1)}  \bigg) \Delta \ln a^{(1)} +\bigg[ \frac{2}{\bar \chi}\left(\frac{\cH' }{\cH^3} +\frac{1}{\cH} \right)-\frac{\cH'' }{\cH^3} +2\left( \frac{\cH' }{\cH^2} \right)^2 + \frac{\cH' }{\cH^2} - b_e +  \frac{\ud \ln b_e}{\ud  \ln \bar a}  - \frac{2}{\bar \chi^2 \cH^2}\bigg] \left[ \Delta \ln a^{(1)} \right]^2 \nonumber \\  
&&+\left( \frac{2}{\cH} \p_\|^3 v -\frac{2}{\cH} \p_\| \Phi {'} -\frac{8}{\bar \chi}\Phi  +2\p_{\|}\Phi   -2\p_{\|}\delta_g   -  \frac{4}{\bar \chi}  \kappa^{(1)}  - \frac{2}{\bar \chi^2} T^{(1)} \right) T^{(1)}  + 4 \bigg[  4 \Phi  + \frac{\cH' }{\cH^3}\p_\|^2 v -\frac{\cH' }{\cH^3}\Phi {'} +\frac{1}{\cH} \Phi {'}    +  \frac{1}{\cH} \p_\|^2 v  \nonumber \\ 
&&    +  \frac{1}{\cH^2} \p_\|^2 \Phi  + \frac{1}{\cH^2} \p_\|^3 v  + \frac{1}{\cH}   \p_\| \Phi - \frac{1}{\cH^2}\frac{\ud \,}{\ud \bar \chi}\Phi {'}   -2 I^{(1)}  \bigg] I^{(1)} +8\left( \Phi   -  I^{(1)}  \right)  \int_0^{\bar \chi} \ud \tilde \chi\bigg[  \frac{\tilde \chi}{ \bar \chi} \left( 2 \tilde \p_\| \Phi + \left(\bar \chi-\tilde \chi\right)  \Perp^{mn} \tilde \p_m  \tilde \p_n \Phi  \right)  \bigg]  \nonumber \\
&& -4\bigg[  \int_0^{\bar \chi} \ud \tilde \chi  \frac{\tilde \chi}{ \bar \chi} \left( \Perp_j^i \tilde \p_\| \Phi + \left(\bar \chi-\tilde \chi\right) \Perp^p_j \Perp^{iq} \tilde \p_q  \tilde \p_p\Phi  \right) \bigg]  \bigg[\int_0^{\bar \chi} \ud \tilde \chi \frac{\tilde \chi}{ \bar \chi} \left( \Perp^j_i \tilde \p_\| \Phi + \left(\bar \chi-\tilde \chi\right) \Perp^n_i \Perp^{jm}  \tilde \p_m   \tilde \p_n \Phi   \right)   \bigg]\nonumber \\
 && + 8\int_0^{\bar \chi}  \ud \tilde{\chi}\bigg[   -  \tilde \p_{\perp j} \Phi    S_{\perp}^{j(1)}    -  \Phi  \tilde \p_{\perp m}S_{\perp}^{m(1)}  +  \left( \frac{\ud \Phi}{\ud \tilde \chi}   -  \frac{1}{\tilde\chi} \Phi  \right) \kappa^{(1)}  \bigg]  +  \frac{8}{\bar \chi}\int_0^{\bar \chi}  \ud \tilde{\chi} \bigg[  -  {\Phi}^2  -2 S_{\perp }^{i (1)}S_{\perp }^{j (1) } \delta_{ij}   -  \Phi{'} T^{(1)}         \nonumber \\
 && -2 \Phi \kappa^{(1)}   + 2  \tilde \chi  \tilde \p_{\perp i}\Phi  S_\perp^{i(1)} - \tilde \chi \tilde \p_{\perp i}\Phi  \p_\perp^i T^{(1)}   \bigg] +  \frac{8}{\bar \chi} \int_0^{\bar \chi}  \ud \tilde{\chi} ~ (\bar \chi - \tilde \chi) \bigg[- 2 \tilde \p_{\perp j} \Phi  S_{\perp}^{i(1)}  -  2 \Phi   \tilde \p_{\perp m}S_{\perp}^{m(1)}     
    \nonumber \\
&& + 2  \left( \frac{\ud  \Phi}{\ud \tilde \chi}   -  \frac{1}{\tilde\chi} \Phi  \right) \kappa^{(1)}  \nonumber \bigg] - 2   \bigg[   + \bar \chi  \p_{\perp i}\left(  - \Phi   +\delta_g  \right)-\frac{\bar \chi}{\cH}\p_{\perp i}\bigg( -\Phi {'} +\p_\|^2 v  \bigg) +\frac{2}{\cH} \p_{\perp i}\p_\| v    -\frac{2}{\cH\bar \chi}\p_{\perp i} v \bigg]   \p_\perp^i T^{(1)}\nonumber \\ 
&&- 8 \Phi \kappa^{(1)}    + 4 \bigg[  -\bar \chi  \p_{\perp i} \Phi   +\bar \chi  \p_{\perp i}\delta_g  +\frac{1}{\cH}  \p_{\perp i}   \Delta \ln a^{(1)}   +\frac{\bar \chi}{\cH} \p_{\perp i} \Phi {'} -\frac{\bar \chi}{\cH} \p_{\perp i} \p_\|^2 v       \bigg] S_{\perp }^{i (1)}\nonumber \\
&& -v_{\perp i \, o }  v^{i}_{\perp \, o }  -2 {v _{\| o}}^2-8 {\Phi _{o}}^2-16 \Phi _{ o}v _{\| o} + \left( \Phi _{o} -  v _{\| o} \right)\bigg\{  8 \Phi +  4\bar \chi  \frac{\ud \Phi}{\ud \bar \chi}  - 8 I^{(1)}   + \frac{8}{\bar \chi} T^{(1)}              
 +2\left(\frac{\cH' }{\cH^3} + \frac{1}{\cH}\right) \p_\|^2 v   \nonumber \\
  & & + 24 \int_0^{\bar \chi}   \frac{\ud \tilde{\chi}}{\tilde \chi} \Phi   - 4\int_0^{\bar \chi} \ud \tilde \chi\bigg[   \frac{\tilde \chi}{ \bar \chi} \left(2\tilde \p_\|\Phi  + \left(\bar \chi-\tilde \chi\right)\Perp^{mn} \tilde \p_m \tilde \p_n  \Phi  \right) \bigg]  +2\left(-\frac{\cH' }{\cH^3}+ \frac{1}{\cH}\right)  \Phi {'}-4\bar \chi \frac{\ud}{\ud \bar \chi}\Phi + \frac{2}{\cH}  \p_\| \Phi   \nonumber \\
  & & + \frac{2}{\cH^2} \p_\|^2 \Phi + \frac{2}{\cH^2}\p_\|^3 v  - \frac{2}{\cH^2}\frac{\ud \,}{\ud \bar \chi}\Phi {'}\bigg\} +4  \Phi _{o} \bigg\{2\left(\Phi   -  I^{(1)}  \right)-  \int_0^{\bar \chi} \ud \tilde \chi\bigg[   \frac{\tilde \chi}{ \bar \chi} \bigg(2\tilde \p_\|\Phi  + \left(\bar \chi-\tilde \chi\right)\Perp^{mn} \tilde \p_m \tilde \p_n  \Phi  \bigg) \bigg]  \nonumber \\
 && -\frac{2}{\bar \chi} T^{(1)}-6 \int_0^{\bar \chi}   \frac{\ud \tilde{\chi}}{\tilde \chi} \Phi \bigg\}  -2v_{\perp i \, o }\bigg[ \bar \chi \p^i_{\perp}\left( -  \Phi  + \delta_g  \right) +  \frac{1}{\cH}  \p^i_{\perp}   \Delta \ln a^{(1)}    -\frac{\bar \chi}{\cH}\p_{\perp}^i  \left(-\Phi {'} +\p_\|^2v  \right)  -2S^{i (1)}_{\perp} \bigg], %\nonumber \\
 \end{eqnarray}

where 
\begin{eqnarray}
\label{Poiss-Deltalna-2}
 \Delta\ln a^{(2)}&=&- \Phi^{(2)}+ \p_\| v^{(2)}+ \hat v^{(2)}_\| +  3  {\Phi}^2  -  \left( \p_\| v \right)^2+ \p_{\perp i} v \,  \p^i_{\perp} v  -2  \p_\| v \,  \Phi  - \frac{2}{\cH}\left( \Phi  -  \p_\| v \right) \left(  \Phi {'} - \p_\|^2 v  \right)  \nonumber \\
&& - 4 \bigg[ 3 \Phi  +  \frac{1}{\cH}  \p_\|^2 v   - \frac{1}{\cH}  \Phi {'} -2\bar \chi  \p_\|  \Phi \bigg] I^{(1)}   +2  \;  \p_\| \left( \Phi -  \p_\| v \right) T^{(1)}  +8   \p_\|  \Phi \int_0^{\bar \chi} \ud \tilde \chi \tilde \chi\Phi {'}  +   4 \bar \chi \p_{\perp i}\left(\Phi  +   \p_\| v  \right)   S_{\perp}^{i (1)} \nonumber \\
         && +   8\Phi   I^{(1)} +8\Phi  \kappa^{(1)}  +  4  \Phi{'} T^{(1)}    - 8 \bar \chi \p_\|  \Phi I^{(1)} - 8 \p_\|  \Phi \int_0^{\bar \chi} \ud \tilde \chi ~  \tilde \chi  \Phi{'}    -8 \bar \chi \p_{\perp i}\Phi   S_\perp^{i(1)} +4\bar \chi \p_{\perp i}\Phi   \p_\perp^i T^{(1)}  \nonumber\\
    &&  +4 \int_0^{\bar \chi}  \ud \tilde{\chi} \Bigg[   \Phi{''}  T^{(1)}  + 2 \Phi \Phi{'}  +  2 \Phi{'}  I^{(1)} + 2 S_{\perp}^{i(1)}  \tilde \p_{\perp i}  \Phi    + 2 \Phi    \tilde \p_{\perp j}S_{\perp}^{j(1)}       - 2 \bigg( \frac{\ud}{\ud \tilde \chi} \Phi    -  \frac{1}{\tilde\chi}  \Phi  \bigg) \kappa^{(1)}    - 2  \tilde \chi  \tilde \p_{\perp i} \Phi{'}    S_\perp^{i(1)}  \nonumber\\
&&+ \tilde \chi  \tilde \p_{\perp i} \Phi{'}   \p_\perp^i T^{(1)} \Bigg]     -2 \bigg[\bar \chi \p_{\perp i}\left(\Phi  +   \p_\| v  \right)   - \p_{\perp i} v \bigg]  \p_{\perp}^i T^{(1)}+ 2I^{(2)} + 8  \left(I^{(1)}\right)^2+4\delta_{ij} S_{\perp }^{i (1)} S_{\perp }^{j (1)}\nonumber \\
&& + \Phi^{(2)}_{o}- v^{(2)}_{\| o} +  8\Phi_o v_{\| \, o} - \Phi_o^2+  v _{k\, o} v^{k  }_o  + 2\left(\Phi_o-v _{\| o}\right) \bigg(- \Phi - \frac{1}{\cH}  \p_\|^2 v  + \frac{1}{\cH}  \Phi {'}    + 2 I^{(1)}  \bigg)  + 8 v_{\| \, o}    \int_0^{\bar \chi}   \frac{\ud \tilde{\chi}}{\tilde \chi}  \Phi   \nonumber \\
&&  - 2 v_{\perp i \, o}  \left[ - \bar \chi \, \p_{\perp}^i \left(\Phi  +   \p_\| v  \right)+2 \bar \chi \p^i_{\perp }  I^{(1)}  \right].
\end{eqnarray}
 
At second order the lensing convergence term \eqref{kappa-n} is
\begin{eqnarray}
 \label{Poiss-kappa-2}
 \kappa^{(2)}&=& \frac{1}{2}  \int_0^{\bar \chi} \ud \tilde \chi  \left(\bar \chi-\tilde \chi\right) \frac{\tilde \chi}{ \bar \chi}   \tilde \nabla^2_\perp \left( \Phi^{(2)} + 2 \omega^{(2)}_{\| }+\Psi^{(2)} - \frac{1}{2} \hat h^{(2)}_{\| } \right)  + \frac{1}{2}  \int_0^{\bar \chi} \ud \tilde \chi \bigg(-2  \tilde \p_\perp^i \omega_i^{ (2)} + \frac{4}{\tilde\chi} \omega_\|^{ (2)}+ \Perp^{ij} n^k  \tilde \p_i  \hat h_{jk}^{ (2)} \nonumber \\
 && - \frac{3}{\tilde \chi}  \hat h_\|^{ (2)}\bigg)
 -2\bigg( 2 \bar\chi I^{(1)} +2\int_0^{\bar \chi} \ud \tilde \chi \tilde \chi \Phi {'}  +T^{(1)}   +  \frac{1}{\cH}\Delta \ln a^{(1)}   \bigg)   \int_0^{\bar \chi} \ud \tilde \chi \bigg( \frac{\tilde \chi}{ \bar \chi} \,  \tilde \nabla^2_\perp \Phi  \bigg) -2S_{\perp}^{i (1)}   \bigg[ - \p_{\perp i } T^{(1)}  \nonumber \\
 &&   -  \frac{1}{\cH} \p_{\perp i}  \Delta \ln a^{(1)}   -2 \bar \chi \p_{\perp i}  I^{(1)} -  2\p_{\perp i} \int_0^{\bar \chi} \ud \tilde \chi ~  \tilde \chi  \Phi{'} \bigg]  
  + 4 \int_0^{\bar \chi} \ud \tilde \chi \, \frac{\tilde \chi}{ \bar \chi}  \bigg[  -   \tilde \p_{\perp j} \Phi  S_{\perp}^{j (1)}+ \frac{2}{\tilde \chi}   \Phi  S_{\|}^{(1)}  -  \Phi   \tilde \p_{\perp m} S^{m (1)}\bigg]   \nonumber \\ 
&& - 4\int_0^{\bar \chi} \ud \tilde \chi \bigg\{ \left(\bar \chi-\tilde \chi\right)\frac{\tilde \chi}{ \bar \chi}  \bigg[\left(  \tilde \p_{\perp i} \Phi - 2  \tilde \p_{\perp i}I^{(1)}  \right)\tilde\p^i_\perp  \Phi +\left( \Phi - 2 I^{(1)}\right)   \tilde \nabla^2_\perp \Phi\bigg] \bigg\}   +2 \int_0^{\bar \chi} \ud \tilde{\chi}  \frac{ \tilde \chi }{\bar \chi} \Bigg[  + 2\tilde \chi   \tilde \nabla^2_{\perp} \Phi I^{(1)}\nonumber \\
  &&   + 2  \tilde \nabla^2_{\perp} \Phi \int_0^{\tilde \chi} \ud \tilde{\tilde \chi} \tilde{\tilde \chi}  \Phi {'}   + 2  \tilde \chi \p_{\perp i}  \Phi   \tilde \p_\perp^i I^{(1)} + 2 \p_{\perp i}  \Phi   \tilde \p_{\perp i} \int_0^{\tilde \chi} \ud \tilde{\tilde \chi} \tilde{\tilde \chi}  \Phi {'}  + 2  \tilde \p_{\perp i} \Phi   S_\perp^{i(1)}  -  \tilde \p_{\perp i} \Phi  \p_\perp^i T^{(1)}  - \frac{2}{\tilde \chi}\Phi  \kappa^{(1)}  \Bigg] \nonumber\\
&&    +2 \int_0^{\bar \chi} \ud \tilde{\chi} ~ (\bar \chi - \tilde \chi) \frac{ \tilde \chi }{\bar \chi} \Bigg[ - \tilde \nabla^2_{\perp} \Phi  ~ T^{(1)}    -   \tilde \p_{\perp }^i    \Phi{'}  \tilde \p_{\perp i} T^{(1)}  -2 I^{(1)}   \tilde \nabla^2_{\perp}\Phi    -2 \p_{\perp i} \Phi ~ \tilde \p^i_{\perp} I^{(1)}   +  \frac{2}{\tilde\chi}  \bigg( -     \frac{1}{\tilde\chi}\Phi +  \frac{\ud}{\ud \tilde \chi} \Phi\bigg)  \kappa^{(1)}  \nonumber\\
&&  +  \frac{1}{\tilde\chi} \tilde \p_{\perp i}  \Phi ~ S_{\perp}^{i(1)}    - \frac{3}{2 \tilde \chi}  \tilde \p_{\perp i} \Phi   \p_\perp^i T^{(1)}    + \tilde \chi \bigg(  \tilde \p_{\perp i} \tilde \nabla^2_{\perp} \Phi + \frac{1}{\tilde\chi} \tilde \p_{\perp i}  \Phi{'}    \bigg)\left( 2   S_\perp^{i(1)} -  \p_\perp^i T^{(1)}\right)  - \left( \Phi{'}   + \frac{1}{\tilde \chi} \Phi  \right) \tilde \nabla_\perp^2 T^{(1)} \nonumber\\
&& +  \tilde\chi \tilde \p^{(j}_{\perp} \tilde \p^{m)}_{\perp} \Phi \left( 2  \p_{\perp(m}  S_{\perp j)}^{(1)} -   \p_{\perp (m}   \p_{\perp j)}  T^{(1)}\right)  +  2\Phi{'} \tilde \p_{\perp m} S_{\perp}^{m(1)}  \Bigg]  - 2 \omega_{\| o}^{(2)}- v_{\| o}^{(2)}+\frac{3}{4}\hat h_{\|  o}^{ (2)}     + 4 v_{\| \, o} \Phi_o- v _{\| \,o}^2 \nonumber \\
     && + \frac{1}{2} v _{\perp i \, o } v^{i  }_{\perp \, o}     + \left(\Phi _o-v _{\| \, o}\right) \left( - \bar \chi  \nabla^2_{\perp} T^{(1)}    - 2    \bar \chi^2 \nabla^2_{\perp} \int_0^{\bar \chi}   \frac{\ud \tilde{\chi}}{\tilde \chi} \Phi    + 2\kappa^{(1)}\right) +2v_{\| o}  \bigg( 2  I^{(1)} + \kappa^{(1)}   + \frac{2}{\bar \chi} \int_0^{\bar \chi} \ud \tilde \chi \tilde \chi  \Phi {'}  \nonumber \\
    &&+\frac{2}{\bar \chi} T^{(1)}    + \frac{1}{\cH}\frac{1}{\bar \chi}\Delta \ln a^{(1)}  \bigg)  -  v^i_{\perp \, o } \bigg[  + 2  S_\perp^{i(1)} + 2\p_\perp^i T^{(1)}  +\bar \chi^2   \p_{\perp i}  \nabla^2_{\perp}\left( T^{(1)}+2 \bar \chi  \int_0^{\bar \chi}   \frac{\ud \tilde{\chi}}{\tilde \chi} \Phi \right)  +4 \bar \chi     \p_{\perp i}  \int_0^{\bar \chi}   \frac{\ud \tilde{\chi}}{\tilde \chi} \Phi\nonumber \\
      &&    +  \frac{1}{\cH} \p_{\perp i}  \Delta \ln a^{(1)} \bigg]  \;.
  \end{eqnarray}

The second order forms of \eqref{Poiss-iota}--\eqref{Poiss-varsigma} are
\begin{eqnarray}
\label{Poiss-iota2}
I^{(2)}& =& -\frac{1}{2} \int_0^{\bar \chi} \ud \tilde \chi \left(\Phi^{(2)}{'} +2  \omega^{(2)}_{\| }{'} + \Psi^{(2)}{'}- \frac{1}{2} \hat h^{(2)}_{\| }{'} \right),
\\
\label{Poiss-varsigma2}
S_{\perp}^{i(2)} &=& -\frac{1}{2} \int_0^{\bar \chi} \ud \tilde \chi \left[ \tilde\p^i_\perp \left( \Phi^{(2)} +2  \omega^{(2)}_{\| } + \Psi^{(2)}- \frac{1}{2} \hat h^{(2)}_{\| }\right) + \frac{1}{\tilde \chi} \left(-2 \omega^{i (2)}_{\perp }+  n^k \hat h_{kj}^{(2)} \Perp^{ij}  \right)\right], \\
\label{Poiss-s-1}
 T^{(2)} &=& - \int_0^{\bar \chi} \ud \tilde \chi \left(\Phi^{(2)} +2 \omega^{(2)}_{\| }+ \Psi^{(2)}- \frac{1}{2}h^{(2)}_{\| }\right) .
\end{eqnarray}

~\\
\noindent{\bf Weak lensing  shear and rotation terms}\\

We can simplify $\Delta^{(2)}_g$ by explicitly introducing the weak lensing shear $\gamma_{ij}^{(1)}$ and rotation $\vartheta_{ij}^{(1)}$, defined by
\begin{equation}
\label{shear}
\gamma_{ij}^{(1)}=- \p_{\perp (i}   \Delta x_{\perp j)}^{(1)}-\Perp_{ij}\kappa^{(1)}, \quad\quad\quad\quad\quad
\vartheta_{ij}^{(1)}=-\p_{\perp [i}   \Delta x_{\perp j]}^{(1)}.
\end{equation}
These do not contribute to the observed number counts at first order but quadratic products do contribute at second order.

Then \eqref{Poiss-Deltag-2} becomes
\begin{eqnarray}
\label{Poiss-Deltag-2s}
 \Delta_g^{(2)} &=&  \delta_g^{(2)} + \Phi^{(2)}  - 2  \Psi^{(2)} -\frac{1}{2}\hat h_{\|}^{(2)}+  \frac{1}{ \cH} \Psi^{(2)}{'}-  \frac{1}{2 \cH} \hat h^{(2)}_{\| }{'}   -\frac{1}{ \cH }\p_\|^2v^{(2)} -\frac{1}{ \cH }\p_\| \hat v_\|^{(2)}  +  \left( b_e  -   \frac{\cH'}{\cH^2} -\frac{2}{\bar \chi \cH}\right) \, \Delta \ln a^{(2)}  \nonumber \\
 && \nonumber \\
&& - \frac{2}{\bar \chi} T^{(2)}  - 2\kappa^{(2)} + \left(\Delta_g  \right)^2 + \frac{1}{\cH^2}\left(\p_\|^2 v  \right)^2 -6   {\Phi}^2+\left( \p_\| v  \right)^2 +\frac{4}{\bar \chi  \cH} \left( \p_\| v  \right)^2+ \frac{1}{\cH^2}\left( \Phi {'}  \right)^2  
  - \left(\delta_g \right)^2+\frac{2}{\cH}\p_\| v  \Phi {'} \nonumber \\ 
&&  + \frac{2}{ \cH }\Phi  \Phi {'}+2\frac{\cH'}{\cH^3}\Phi \Phi {'}+\frac{4}{ \cH }\p_\| v  \p_\| \Phi  - \frac{4}{\cH} \Phi \p_\|^2 v   - \frac{2}{\cH^2} \Phi \p_\|^3 v   -\frac{2}{\cH}\Phi \p_\| \Phi + \frac{2}{\cH^2}\Phi \frac{\ud \,}{\ud \bar \chi}\Phi {'} - \frac{2}{\cH^2}\p_\| v \frac{\ud \,}{\ud \bar \chi}\Phi {'}   \nonumber \\ 
&&   + \frac{2}{\cH^2} \p_\| v  \p_\|^2 \Phi  -2\frac{\cH'}{\cH^3}\Phi  \p_\|^2 v    +\frac{6}{\cH} \p_\| v \p_\|^2 v  +2\frac{\cH'}{\cH^3} \p_\| v \p_\|^2 v  - \frac{2}{\cH^2}\Phi  \p_\|^2 \Phi   -2\frac{\cH'}{\cH^3} \p_\| v \Phi {'}-  \frac{2}{\cH^2}\p_\|^2 v   \Phi {'}     +\frac{2}{\cH}\p_{\perp i} v \p^i_\perp \Phi  \nonumber \\  
&&   -\frac{2}{ \cH } \p_{\perp i} v   \p_{\perp}^i \p_\|  v  +\frac{2}{ \bar \chi \cH } \p_{\perp i} v    \p_{\perp}^i v   - \p_{\perp i} v   \p_{\perp}^i v  +  \frac{2}{\cH^2}\p_\| v \p_\|^3v  +\frac{2}{ \cH } \p_\| v   \nabla^2_{\perp}  v  + \bigg( - \frac{8}{\bar\chi \cH}   \Phi  -4\frac{1}{\cH} \Phi {'}   - \frac{2}{\cH}\frac{\ud \,}{\ud  \bar \chi} \delta_g   \nonumber \\  
&&  -  \frac{4}{\cH \bar \chi^2}  T^{(1)} - \frac{4}{\bar \chi \cH}   \kappa^{(1)} \bigg) \Delta \ln a^{(1)} +\bigg[ \frac{2}{\bar \chi}\left(\frac{\cH' }{\cH^3} +\frac{1}{\cH} \right)-\frac{\cH'' }{\cH^3} +2\left( \frac{\cH' }{\cH^2} \right)^2 + \frac{\cH' }{\cH^2} - b_e +  \frac{\ud \ln b_e}{\ud  \ln \bar a} - \frac{2}{\bar \chi^2 \cH^2}\bigg]   \nonumber \\  
&&\times\left[ \Delta \ln a^{(1)} \right]^2 +\left( \frac{2}{\cH} \p_\|^3 v -\frac{2}{\cH} \p_\| \Phi {'}  -\frac{8}{\bar \chi}\Phi  +2\p_{\|}\Phi   -2\p_{\|}\delta_g   -  \frac{4}{\bar \chi}  \kappa^{(1)} - \frac{2}{\bar \chi^2} T^{(1)} \right) T^{(1)}  + 4 \bigg[   \frac{\cH' }{\cH^3}\p_\|^2 v -\frac{\cH' }{\cH^3}\Phi {'} \nonumber \\ 
&& +\frac{1}{\cH} \Phi {'}   +  \frac{1}{\cH} \p_\|^2 v  +  \frac{1}{\cH^2} \p_\|^2 \Phi  + \frac{1}{\cH^2} \p_\|^3 v  + \frac{1}{\cH}   \p_\| \Phi  - \frac{1}{\cH^2}\frac{\ud \,}{\ud \bar \chi}\Phi {'}   \bigg] I^{(1)}- 8  \bigg(\bar \chi \frac{\ud}{\ud \bar \chi} \Phi   + \frac{\bar \chi}{\cH} \frac{\ud \,}{\ud \bar \chi} \p_\| \Phi +2 \Phi   \bigg)  \int_0^{\bar \chi} \ud \tilde \chi \left(\frac{\tilde \chi}{\bar \chi}\Phi {'} \right) \nonumber \\  
&&  -2\big|\gamma^{(1)}\big|^2-2\left(\kappa^{(1)}\right)^2+\vartheta_{ij}^{(1)}\vartheta^{ij(1)} - 2   \bigg[   \bar \chi  \p_{\perp i}\left(  - \Phi   +\delta_g  \right) + \frac{\bar \chi}{\cH}\p_{\perp i} \Phi {'} - \frac{\bar \chi}{\cH}\p_{\perp i} \p_\|^2 v   +\frac{2}{\cH} \p_{\perp i}\p_\| v  \ -\frac{2}{\cH\bar \chi}\p_{\perp i} v \bigg]   \p_\perp^i T^{(1)} \nonumber \\ 
&&  + 4 \bigg[      -\bar \chi  \p_{\perp i} \Phi   +\bar \chi  \p_{\perp i}\delta_g  +\frac{1}{\cH} \p_{\perp i}    \Delta \ln a^{(1)}   +\frac{\bar \chi}{\cH} \p_{\perp i} \Phi {'} -\frac{\bar \chi}{\cH} \p_{\perp i} \p_\|^2 v  \bigg] S_{\perp }^{i (1)}+ 8 \left(\frac{2}{\bar \chi}  \Phi+\frac{\ud \Phi}{\ud \bar \chi}     +\frac{1}{\cH} \frac{\ud}{\ud \bar \chi} \p_\| \Phi   \right)   \nonumber \\
 &&\times \int_0^{\bar \chi} \ud \tilde \chi  \tilde \chi  \Phi{'} - 8 \Phi \kappa^{(1)}   +  8 \int_0^{\bar \chi}  \ud \tilde{\chi}\bigg[     -  \tilde \p_{\perp j} \Phi    S_{\perp}^{j(1)}   -  \Phi  \tilde \p_{\perp m}S_{\perp}^{m(1)}  +  \left( \frac{\ud \Phi}{\ud \tilde \chi}   -  \frac{1}{\tilde\chi} \Phi  \right) \kappa^{(1)}  \bigg] +  \frac{8}{\bar \chi}\int_0^{\bar \chi}  \ud \tilde{\chi} \bigg[ -  \Phi{'} T^{(1)}  \nonumber \\
 &&     - 4 \Phi \kappa^{(1)}    + 2  \tilde \chi  \tilde \p_{\perp i}\Phi  S_\perp^{i(1)}  - \tilde \chi \tilde \p_{\perp i}\Phi  \p_\perp^i T^{(1)}  -  {\Phi}^2-2 S_{\perp }^{i (1)}S_{\perp }^{j (1)} \delta_{ij}  \bigg]    +  \frac{8}{\bar \chi} \int_0^{\bar \chi}  \ud \tilde{\chi} ~ (\bar \chi - \tilde \chi) \bigg[   - 2 \tilde \p_{\perp j} \Phi  S_{\perp}^{i(1)} \nonumber \\
 &&  -  2 \Phi   \tilde \p_{\perp m}S_{\perp}^{m(1)}    + 2  \left( \frac{\ud  \Phi}{\ud \tilde \chi}   -  \frac{1}{\tilde\chi} \Phi  \right) \kappa^{(1)}   \bigg] -v_{\perp i \, o }  v^{i}_{\perp \, o } -24 \Phi _{\, o}v _{\| \, o}  - 8   v_{\| \, o}  \left(  \frac{1}{\bar \chi} T^{(1)}+ 3 \int_0^{\bar \chi}   \frac{\ud \tilde{\chi}}{\tilde \chi} \Phi \right) \nonumber \\
&& + \left( \Phi _{\, o} -  v _{\| \, o} \right)\bigg[  2\left(\frac{\cH' }{\cH^3} + \frac{1}{\cH}\right) \p_\|^2 v     +2\left(-\frac{\cH' }{\cH^3}+ \frac{1}{\cH}\right)  \Phi {'} + \frac{2}{\cH}  \p_\| \Phi   + \frac{2}{\cH^2} \p_\|^2 \Phi + \frac{2}{\cH^2}\p_\|^3 v  - \frac{2}{\cH^2}\frac{\ud \,}{\ud \bar \chi}\Phi {'}\bigg]  \nonumber \\
 && -2v_{\perp i \, o }\bigg[ \bar \chi \p^i_{\perp}\left( -  \Phi  + \delta_g  \right) +  \frac{1}{\cH}  \p^i_{\perp}   \Delta \ln a^{(1)}   -\frac{\bar \chi}{\cH}\p_{\perp}^i  \left(-\Phi {'} +\p_\|^2v \right)  -2S^{i (1)}_{\perp} \bigg],  \nonumber \\
 \end{eqnarray}
          \begin{eqnarray} 
   &&  \nonumber \\
     &&     \nonumber \\
 &&  \nonumber \\
   \end{eqnarray}
 where $2|\gamma^{(1)}|^2=\gamma_{ij}^{(1)}\gamma^{ij (1)}$. Explicit expressions for $\gamma_{ij}^{(1)}$  and $\vartheta_{ij}^{(1)}\vartheta^{ij(1)}$ are
  \begin{eqnarray}
\gamma_{ij}^{(1)} &=& - \Perp_{ij} v_{\|\, o}-n_{(j}  v_{\perp i) \, o}  - 2 \int_0^{\bar \chi} \ud \tilde \chi \left[\left(\bar \chi-\tilde \chi\right) \frac{\tilde \chi}{ \bar \chi}  \tilde \p_{\perp (i}  \tilde \p_{\perp j)} \Phi \right]-\Perp_{ij} \kappa^{(1)},\label{gamp}\\
\vartheta_{ij}^{(1)}\vartheta^{ij(1)} &=&   +\frac{1}{2}v_{\perp i o}^{(1)}v_{\perp  o}^{i(1)}+ \frac{2}{ \bar \chi}v_{\perp i \, o} \int_0^{\bar \chi} \ud \tilde \chi\bigg[ \left(\bar \chi-\tilde \chi\right)  \tilde \p^i_{\perp } \Phi\bigg]+\frac{2}{\bar \chi^2}  \int_0^{\bar \chi} \ud \tilde \chi\bigg[ \left(\bar \chi-\tilde \chi\right)  \tilde \p_{\perp i}  \Phi \bigg]  \times \int_0^{\bar \chi} \ud \tilde \chi\bigg[ \left(\bar \chi-\tilde \chi\right)  \tilde \p^i_{\perp}  \Phi \bigg] \;. \nonumber \\\label{varp} 
 \end{eqnarray}

%%%%%%%%%%%%%%%%%%%%%%%%%%%%%%%%%%%%%%%%%%%%%%%%%%%%%%%%%%%%%%%%%%%%%%%%%%%%%%%%%%%%%%%%%%%%%%%%%%%%%%%%%%%%%%%%%%%%%%%%%%%%%%
%%%%%%%%%%%%%%%%%%%%%%%		PRESCRIPTIONS FOR GALAXY BIAS  AT SECOND ORDER  %%%%%%%%%%%%%%%%%%%%%%%%%%%%%%%%%%%%%%%%%%%%%%%%%%%%%%%%%%%%%%%%%%%%%%%%%%%%%%%%%%%%%%%%
%%%%%%%%%%%%%%%%%%%%%%%%%%%%%%%%%%%%%%%%%%%%%%%%%%%%%%%%%%%%%%%
~\\
\noindent{\bf Galaxy bias}\\

Fluctuations of galaxy number density are related to the underlying matter density fluctuation $\delta_m$ on cosmological scales by a local bias.
In order to define this correctly,  we need to choose an appropriate frame where the baryon velocity perturbation vanishes. 
The standard assumption at first order is that the baryon velocity is equal to the CDM velocity on large scales, i.e. well above the nonlinear scale. Since we are dealing with large scales, it seems reasonable to extend the standard assumption to second order.
Then  the baryon rest frame coincides with the CDM rest frame and   in $\Lambda$CDM, this rest frame is defined up to second order by the comoving-synchronous  gauge (S)  \cite{Matarrese:1997ay, Wands:2009ex, Bartolo:2010rw, Bartolo:2010ec, Bruni:2011ta, Bruni:2013qta}. In this gauge, the galaxy and matter overdensities are gauge invariant \cite{Kodama:1985bj}. 
The S-gauge is defined by the conditions  $g_{00}=-1$,    $g_{0i}=0$ and $v^{i}=0$. Then
 \begin{eqnarray} 
 \label{Comoving-ortnogonal_metric}
 \ud s^2 = a(\eta)^2\left\{-\ud\eta^2+\left[\delta_{ij} -2 \psi \delta_{ij} + \left(\p_i\p_j-\frac{1}{3} \delta_{ij}\nabla^2\right) \xi+\frac{1}{2}h_{ij {\rm S}}^{(2)}\right]\ud x^i\ud x^j\right\},
\end{eqnarray}
where $h^{(2)}_{ij {\rm S}} =- 2 \psi^{(2)} \delta_{ij} + F_{ij  {\rm S}}^{(2)}$, with $F^{(2)}_{ij   {\rm S}}= (\p_i\p_j- \delta_{ij}\nabla^2/3) \xi^{(2)}+\p_i \hat \xi^{(2)}_j+ \p_j \hat \xi^{(2)}_i+\hat h^{(2)}_{ij}$, $\p_i \hat \xi^{i(2)}=\p_i \hat h^{ij(2)}=0$. 
%Here, for simplicity, we neglect vector and tensor perturbations at  first order, i.e.  $\hat \xi_j=  \hat h_{ij}=0$\;.
%Here $D^{(n)}$ and $F^{(n)}$ are a scalar, $\hat F^{(n)}_i$ is a solenoidal vector field,   $\p^i \hat h^{(n)}_{ij}=\hat h_i^{i(n)}=0$.

In order to obtain the galaxy fractional number overdensity  $\delta_{g {\rm S}}$, we transform the metric perturbations from the Poisson  to  comoving-synchronous  gauge. We find that 
 \begin{eqnarray} 
 \label{dg1}
\delta_{g }&=&\delta_{g \, {\rm S}}- b_e \cH v+ 3 \cH v ,\\
 \label{dg2}
\delta_{g }^{(2)} &=&  \delta_{g  {\rm S}}^{(2)}- b_e \cH v^{(2)}+ 3 \cH v^{(2)} + \left( b_e \cH'-3 \cH' + \cH^2  \frac{\ud  b_e}{\ud  \ln \bar a} + b_e^2 \cH^2  -6  b_e  \cH^2 + 9 \cH^2 \right) v^2 + \cH b_e  v  {v}' - 3 \cH   v  {v}' \nonumber\\
&&  -2\cH b_e  v \delta_{g  {\rm S}}   + 6 \cH  v \delta_{g  {\rm S}} - 2  v {\delta_{g{\rm S}}}' - \frac{1}{2} \p^i \xi \left(- b_e \cH \p_i v+ 3 \cH \p_i v + 2 \p_i \delta_{g  {\rm S}} \right) - \left(b_e-3\right) \cH  \nabla^{-2}\bigg( v \nabla^2 {v}' - {v}' \nabla^2 v  \nonumber \\
&&- 6 \p_i \Phi \p^i v - 6 \Phi \nabla^2 v  + \frac{1}{2} \p_i \xi \p^i \nabla^2 v + \frac{1}{2} \p_i v \p^i \nabla^2 \xi + \p_i \p_j \xi \p^i \p^j v\bigg).
\end{eqnarray} 
%where
%\begin{eqnarray} 
%\Xi &=&+ v \nabla^2 {v}' - {v}' \nabla^2 v - 6 \p_i \Phi \p^i v - 2 \Phi \nabla^2 v - 4 \Phi\nabla^2 v  \nonumber\\
%&& + \frac{1}{2} \p_i \xi \p^i \nabla^2 v + \frac{1}{2} \p_i v \p^i \nabla^2 \xi + \p_i \p_j \xi \p^i \p^j v \;.
%\end{eqnarray} 
Note the useful relation $v={\xi}'/2$.

Then the scale-independent  bias at first and at second order (down to mildly nonlinear scales) is given by\footnote{A typo in this equation has been corrected.}
%\be
%\label{bias}
% \delta_{g {\rm S}} = b_1 (\eta) \delta_{m  {\rm S}},~~~ \delta_{g  {\rm S}}^{(2)}  = b_2 (\eta)  \delta_{m  {\rm S}}^{(2)}.
% \ee
 \begin{equation}
\label{bias}
\delta_{g {\rm S}}^{(1)}+\frac{1}{2} \delta_{g  {\rm S}}^{(2)} = b_{1}^L \delta_{m  {\rm S}}^{(1)} +\frac{1}{2} b_{1}^L  \delta_{m  {\rm S}}^{(2)} +\frac{1}{2} b_{2}^L \big( \delta_{m  {\rm S}}^{(1)} \big)^2\;. \nonumber
 \end{equation}
 
Expressions \eqref{dg1}--\eqref{bias} can then be substituted into   \eqref{Poiss-Deltag-2}, thus incorporating the bias correctly.

%%%%%%%%%%%%%%%%%%%%%%%%%%%%%%%%%%%%%%%%%%%%%%%%%%%%%%%%%%%%%%%%%%%%%%%%%%%%%%%%%%%%%%%%%%%%%%%%%%%%%%%%%%%%%%%%%%%%%%%%%%%%%%
%%%%%%%%%%%%%%%%%%%%%%%		CONCLUSIONS	%%%%%%%%%%%%%%%%%%%%%%%%%%%%%%%%%%%%%%%%%%%%%%%%%%%%%%%%%%%%%%%%%%%%%%%%%%%%%%%%%%%%%%%%
%%%%%%%%%%%%%%%%%%%%%%%%%%%%%%%%%%%%%%%%%%%%%%%%%%%%%%%%%%%%%%%

~\\
~\\
\noindent{\bf Conclusions}\\

In this letter, we have for the first time given the observed galaxy counts to second order in redshift space on cosmological scales  for a $\Lambda$CDM model, including all general relativistic effects. This is given by \eqref{Poiss-Deltag-2}, and by \eqref{Poiss-Deltag-2s} when we make explicit the lensing shear and rotation contribution.
 
Our result allows for an investigation of  whether general relativistic effects are measurable beyond the linear approximation in the mildly nonlinear regime in future surveys. The second-order effects, especially those involving integrals along the line of sight, may make a non-negligible contribution to the observed number counts which in turn could be important for precision cosmology with galaxy surveys. (Compare related work by \cite{Umeh:2012pn,BenDayan:2012ct,Ben-Dayan:2014swa} on second-order corrections to cosmological distances.)

We have carefully treated the scale-independent galaxy bias up to second order using the comoving-synchronous gauge, in order to correctly incorporate bias in the galaxy overdensity. 
Our results will in particular be important for an accurate analysis of the `contamination' of primordial non-Gaussianity by relativistic projection effects \cite{Bertacca:2014n}.

\[\]{\bf Acknowledgments:}\\
We thank Enea di Dio, Ruth Durrer, Giovanni Marozzi, Obinna Umeh for helpful discussions.
DB and RM  are supported by the South African Square Kilometre
Array Project. RM acknowledges support from the UK Science \& Technology Facilities Council (grant ST/K0090X/1). RM and CC are supported by the South African National Research Foundation. We thank Ruth Durrer for alerting us to the possibility of an error in our results.

\end{document}